\documentclass[twocolumn,superscriptaddress,prl,floatfix,amsmath,amssymb]{revtex4-1}
\usepackage[]{graphicx}
\usepackage[usenames,dvipsnames]{color}
\usepackage{soul}
\usepackage{bm}
\usepackage{float}
\usepackage{mathtools}
\usepackage{physics}

\usepackage[utf8]{inputenc}
\usepackage{mathtools}
\usepackage{bpchem}
\usepackage{physics}
\usepackage{bm}
\usepackage{bbm}
\usepackage{datetime}
\usepackage{graphicx}
\usepackage[usenames,dvipsnames,table]{xcolor}
\usepackage{tikz}
\usetikzlibrary{calc}
\usepackage{pgfplots}
\usepackage{pgfplotstable}
\usepackage{asymptote}
\usepackage[caption=false]{subfig}
\usepackage{csquotes}

\usepackage{datetime}
\usepackage{marginnote}
\usepackage{siunitx}

\usepackage{blindtext}

\usepackage{hyperref}
\usepackage{cleveref}

\newcommand{\vecsigma}{\boldsymbol{\sigma}}

\let\oldhat\hat
\renewcommand{\hat}[1]{\oldhat{\mathbf{#1}}}

\crefformat{equation}{#2Eq.~(#1)#3}
\crefformat{figure}{#2Fig.~(#1)#3}
\crefformat{table}{#2Tab.~(#1)#3}

\begin{document}

\title{Unconventional superconductivity in a doped quantum spin Hall
  insulator}

\author{Xianxin Wu}
\affiliation{Institut f\"{u}r Theoretische Physik und Astrophysik, Universit\"{a}t W\"{u}rzburg, Am Hubland Campus S\"{u}d, W\"{u}rzburg 97074, Germany}

\author{Mario Fink}
\affiliation{Institut f\"{u}r Theoretische Physik und Astrophysik, Universit\"{a}t W\"{u}rzburg, Am Hubland Campus S\"{u}d, W\"{u}rzburg 97074, Germany}

\author{Werner Hanke}
\affiliation{Institut f\"{u}r Theoretische Physik und Astrophysik, Universit\"{a}t W\"{u}rzburg, Am Hubland Campus S\"{u}d, W\"{u}rzburg 97074, Germany}

\author{Ronny Thomale}
\affiliation{Institut f\"{u}r Theoretische Physik und Astrophysik, Universit\"{a}t W\"{u}rzburg, Am Hubland Campus S\"{u}d, W\"{u}rzburg 97074, Germany}

\author{Domenico Di Sante}
\affiliation{Institut f\"{u}r Theoretische Physik und Astrophysik, Universit\"{a}t W\"{u}rzburg, Am Hubland Campus S\"{u}d, W\"{u}rzburg 97074, Germany}\email{domenico.disante@physik.uni-wuerzburg.de}

\date{\today}

\begin{abstract}
A monolayer of jacutingaite (Pt$_2$HgSe$_3$) has recently been
identified as a novel quantum spin Hall insulator.
By first-principles calculations, we study its Fermiology
in the doped regime and unveil a type-I and type-II van
Hove singularity for hole and electron doping,
respectively. We find that the common link between the propensity for
a topological band gap at pristine filling and unconventional
superconductivity at finite doping roots in the longer ranged
hybridization integrals on the honeycomb lattice.  In a
combined effort of random phase
approximation and functional renormalization group, we find chiral
$d$-wave order for the type-I and odd-parity $f$-wave order for the
type-II regime. 
\end{abstract}

\maketitle

{\it Introduction --} Quantum spin Hall effect and unconventional
superconductivity are among the most intensely studied fields of
contemporary condensed matter research~\cite{KM1,BHZ,RevModPhys.63.239}. At a superficial level, both topical
areas do not seem to be particularly intertwined: in the pursuit of a
quantum spin Hall insulator with most preferable properties, essential
parameters of optimization include spin-orbit coupling and other
single-particle properties to enhance the topological
bulk gap; in order to accomplish a high-$T_c$ unconventional
superconductor as a quantum many-body state of matter, tuning the electronic interaction strength and
profile appears as the most relevant guiding principle. 

Still, superconductivity and topological band insulators or
semimetals have previously faced each other in several contexts. Most
prominently, this holds for the principal topological classification of single-particle scenarios where the emergent particle-hole
symmetry in superconductors plays a pivotal role~\cite{Schnyder}, and for the case
of superconducting proximity effect imposed on a topologically
non-trivial band structure~\cite{FuKane,PhysRevLett.104.040502,PhysRevLett.105.077001,PhysRevLett.105.177002}. All these instances,
however, do not
include the joint avenue of a superconductor and a
topological band insulator in the same material at only different
doping. Ideally, such a setting might allow for the synthesis of a high-quality domain boundary between
a superconductor and a topological insulator with identical lattice structures,
under the assumption that it were possible to impose distinct gating in both domains. 
Until today, there are only few reports of materials that are believed to
be both  topological insulators and superconductors. Half-Heusler
semimetals~\cite{RPdBi}, Cu-doped Bi$_2$Se$_3$~\cite{CuBi2Se3}, doped BaBiO$_3$~\cite{Yan2013,Li2015}, and 
monolayer WTe$_2$~\cite{Fatemi} are such remarkable exceptions, where a
conventional, i.e., phonon-driven mechanism for superconductivity is
likely to dominate.

\begin{figure}[t]
\centering
\includegraphics[scale=0.3,angle=0,clip=true]{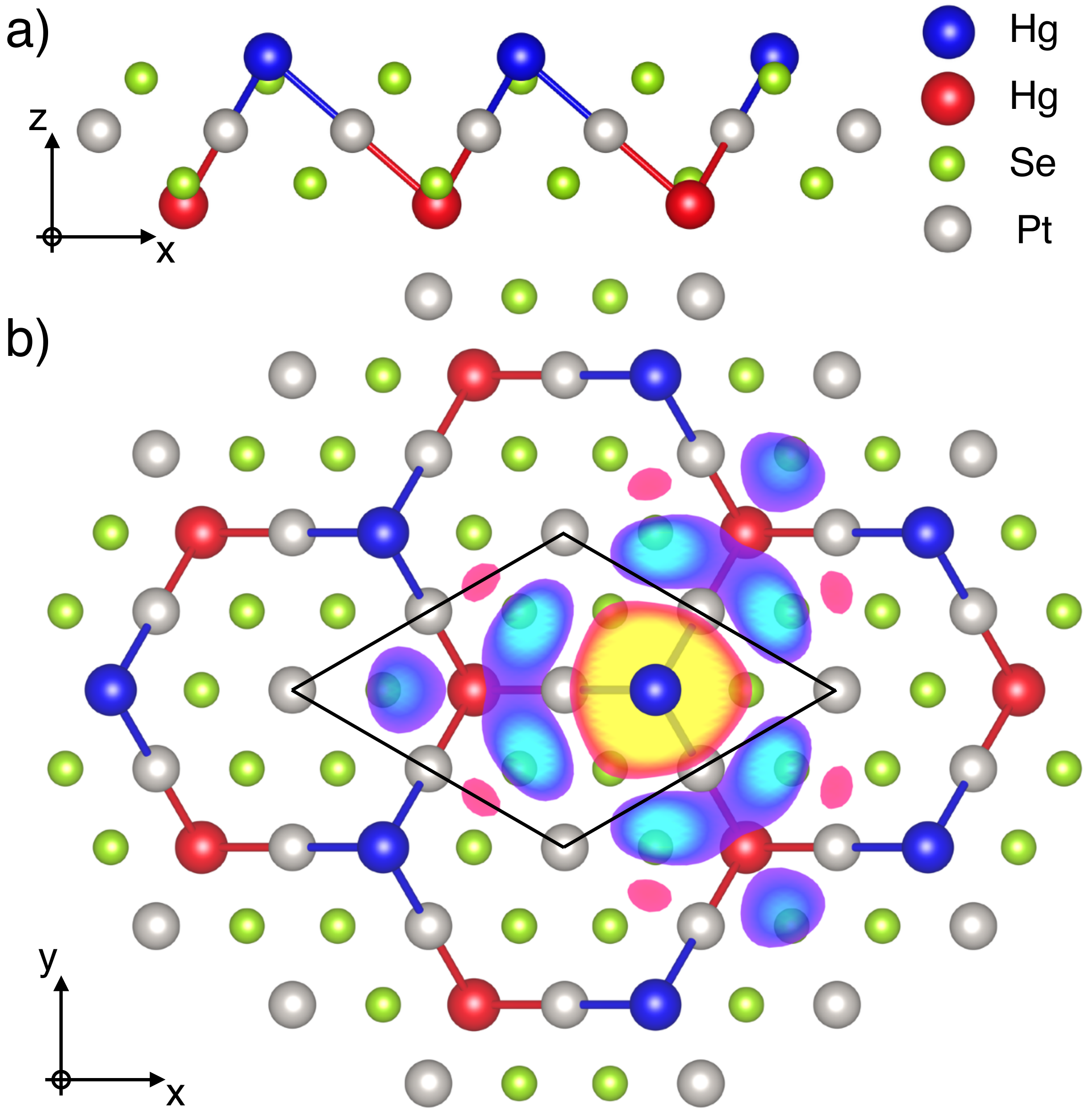}
\caption{(Color online) Side a) and top b) views of the Pt$_2$HgSe$_3$ crystal structure.
In b), a representation of the Wannier function whose in-plane center of mass coincides with the top Hg is shown.
Yellow and blue refer to positive and negative values of the Wannier function. The Wannier
function originating from the bottom Hg can be obtained by inversion.}
\label{fig1}
\end{figure}

In this Letter, we propose a monolayer of jacutingaite
(Pt$_2$HgSe$_3$) to host, besides a quantum spin Hall insulator at pristine
filling~\cite{Marrazzo}, different phases of unconventional superconductivity for
finite hole and electron doping. The central overarching motif that
enables both the realization of quantum spin Hall effect and
unconventional superconductivity is a specific longer ranged
hybridization profile which roots in the extended Wannier functions
of jacutingaite (Fig.~\ref{fig1}). The multi-orbital composition and the honeycomb monolayer
buckling  conspire to yield an effective tight-binding description
which not only provides for a large topological band gap, but also
gives rise to van Hove singularities (vHs) close by pristine filling, with type-I profile
for hole and type-II profile for electron doping. For type-I, the saddle points locate at the time-reversal
invariant momenta (TRIM) {\bf M} of the hexagonal lattice. For type-II,
 the saddle points appear along the {\bf K}$-${\bf M}
lines in the Brillouin zone, and hence do not coincide with
TRIMs~\cite{Yao,vHsII}. As such, while the van Hove induced enhancement of
Fermi level density of states promotes unconventional
superconductivity in general, the nature of the unconventional superconducting state
sensitively depends on the type-I vs. type-II regime, which
we analyze through random phase approximation (RPA)
and functional renormalization group (FRG). We find a $d$-wave instability
for the type-I setting which yields spontaneous time-reversal
symmetry breaking according to a chiral $d$-wave state. For the type-II
setting, the ferromagnetic fluctuations dominate and promote an
odd-parity $f$-wave state.


\begin{figure*}[t]
\centering
\includegraphics[width=\textwidth,angle=0,clip=true]{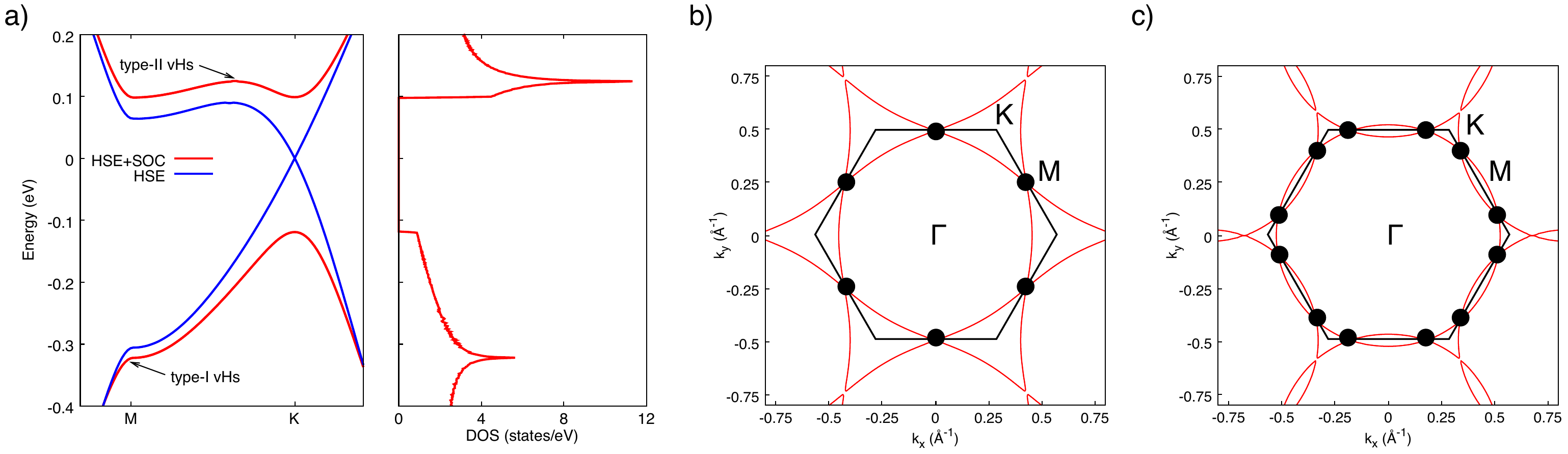}
\caption{(Color online) a) Low energy bandstructure of Pt$_2$HgSe$_3$ at the HSE(+SOC) level of accuracy.
The right panel shows the density of states (DOS) for the calculation including SOC. b-c) Fermi surface at the type-I
and type-II vHs, respectively. The black dots highlight the saddle points where the DOS diverges logarithmically.}
\label{fig2}
\end{figure*}

{\it Effective model --} Monolayer jacutingaite crystallizes in the
spacegroup $P\bar{3}m1$ (164), where the Hg atoms form a buckled
honeycomb lattice surrounded by triangles of Pt and Se
(Fig.~\ref{fig1}a). As first pointed out by Marrazzo et
al.~\cite{Marrazzo}, the low-energy bandstructure of jacutingaite can
be reduced to an effective tight-binding description that shares several
terms with the Kane-Mele model for a quantum spin Hall insulator in
graphene~\cite{KM1}. Anticipating its relevance for jacutingaite at
finite doping, we further add hybridization integrals up to $4$th
nearest neighbor which yields
\begin{eqnarray}
\label{eq1}
H_0^{\text{J}}&=&\sum_{n=1}^4t_n\sum_{\langle ij\rangle_n} c_i^\dagger c_j
+ i\lambda_{\text{SO}} \sum_{\langle ij \rangle_2}\nu_{ij} c_i^\dagger \sigma^z c_j \nonumber \\
&&+ i\lambda_{\text{R}}\sum_{\langle ij\rangle_2} \mu_{ij} c_i^\dagger
(\vecsigma\times\hat{\bf d}_{ij})_z c_j.
\end{eqnarray}
Such a long range hybridization character originates from the delocalized nature of Hg $6s$ and Pt $5d$ orbitals,
which mix to form the hermaphrodite Wannier functions shown in Fig.~\ref{fig1}b.


The
parameters $t_n$, $\lambda_{\text{SO}}$, and $\lambda_{\text{R}}$ are
real, where $t_n$ denotes the $n$th nearest neighbor hopping,
$\lambda_{\text{SO}}$ is the
spin-orbit coupling (SOC) induced second nearest neighbor hopping,
and $\lambda_{\text{R}}$ describes the Rashba SOC.
The absence of a Semenoff mass is due to the centrosymmetric structure
of jacutingaite, i.e., the two Wannier functions
composing the low-energy model have the same on-site energy.
The imaginary parts of nearest and 3rd nearest neighbor hopping vanish due to
mirror (with respect to the Hg-Hg bonds) and time-reversal symmetries.  
Note that $t_2$ connects
equal sublattices on the honeycomb lattice, and as such breaks the chiral
symmetry.
The chiral symmetry operator is $S=\sum_ic^\dag_{is}\tau_{z,ss'}c_{is'}$
where $s(s')$ represents the sublattice index. Since $S^{-1}H_0^{\text{J}}S \ne -H_0^{\text{J}}$ when a real $t_2$ hopping is included,
the resulting energy spectrum ceases to be chiral symmetric in
the presence of finite $t_2$.
The parameters extracted by projecting the density functional theory Hamiltonian
onto a set of maximally localized Wannier orbitals at different levels of sophistication~\cite{Supp}, as shown
in Fig.~\ref{fig2}a for the HSE(+SOC) cases, are summarized in Table~\ref{tab1}.

The topological bulk gap yields $E_g\sim 6\sqrt{3}\lambda_{\text{SO}}$. Note
that $\lambda_{\text{SO}}$ in a Kane-Mele single-orbital scenario describes
an effective SOC which,
starting from the local atomic term, also considers a
downscaling due to the higher-order perturbative effect via second nearest
neighbor hybridization. In a QSH material candidate such as graphene,
this leads to a significant reduction of $\lambda_{\text{SO}}$ because the
longer range hybridization is small~\cite{Liu}, whereas for
jacutingaite, this rescaling is much weaker, combined with the enhanced
atomic SOC of Hg in comparison to C.  An alternative
path to enhance $E_g$ is to realize a two-orbital model per site, as such allowing
for local atomic SOC to affect the low-energy effective model and to
avoid the rescaling due to longer range hybridization. This is
accomplished for bismuthene on SiC~\cite{reis2017bismuthene,PhysRevB.98.165146}. 

\begin{table}[!b]
\centering
\caption{Model parameters extracted by projecting the low-energy states onto two hermaphrodite Wannier orbitals,
that map onto each other under inversion. One of the two is shown in Fig. \ref{fig1}b). All the parameters are given in meV.}
\label{tab1}
\begin{tabular}{p{1.75cm}p{1.0cm}p{1.0cm}p{1.0cm}p{1.0cm}p{1.0cm}p{1.0cm}}
\hline\hline
  & $t_1$ &  $t_2$ &  $\lambda_{\text{SO}}$  &  $\lambda_{\text{R}}$  &   $t_3$    & $t_4$ \\
\hline
PBE+SOC &  168  &  -25  &   18 &  28  &  11  &  -16   \\
HSE+SOC &  265  &  -28  &   21 &  27  & -0.4 &  -28   \\
PBE     &  178  &  -31  &   $-$&  $-$ &  16  &  -24   \\
HSE     &  267  &  -35  &   $-$&  $-$ &  4   &  -32   \\
\hline
\hline
\end{tabular}
\end{table}

Aside from the large gap, further relevant aspects of the resulting band structure are
 visible as we analyze the precise dispersion of the bands in the
 two-dimensional Brillouin zone. By looking at the DOS
in Fig. \ref{fig2}a, vHs peaks arise at the {\bf M} point for the
valence band and along the {\bf K}$-${\bf M} line for the
conduction band, respectively. The Fermi surface at the former vHs (Fig. \ref{fig2}b) shows a hexagonal
profile, with triangular hole pockets around the {\bf K} points.
The Fermi surface at the latter vHs (Fig. \ref{fig2}c), on the
other hand, shows a different shape, with small pockets touching at the saddle points.
We refer to this vHs as type-II \cite{Yao,vHsII}, to distinguish it from the type-I vHs
where the saddle points locate at TRIMs.
A necessary condition for a coexistence of both type-I and type-II vHs in jacutingaite
is a sizable real hopping parameter $t_2$ in~\eqref{eq1}. This
contribution is not contained in the Kane-Mele model~\cite{KM1}, but
indispensable to account for a realistic setting
such as the buckled honeycomb lattice of monolayer jacutingaite. 
To reproduce
the band dispersion given by first-principles calculations, and in
particular to obtain the type-II vHs we find in jacutingaite, longer
range hoppings $t_3$ and $t_4$ need to be taken into consideration (Table~\ref{tab1}).
While those new terms do not change the principal topological nature
of the bulk band gap~\cite{Marrazzo}, they are of primary importance
for an accurate study of pairing states nearby van Hove filling~\cite{Supp}.

\begin{figure}[!h]
\centering
\includegraphics[scale=0.4,angle=0,clip=true]{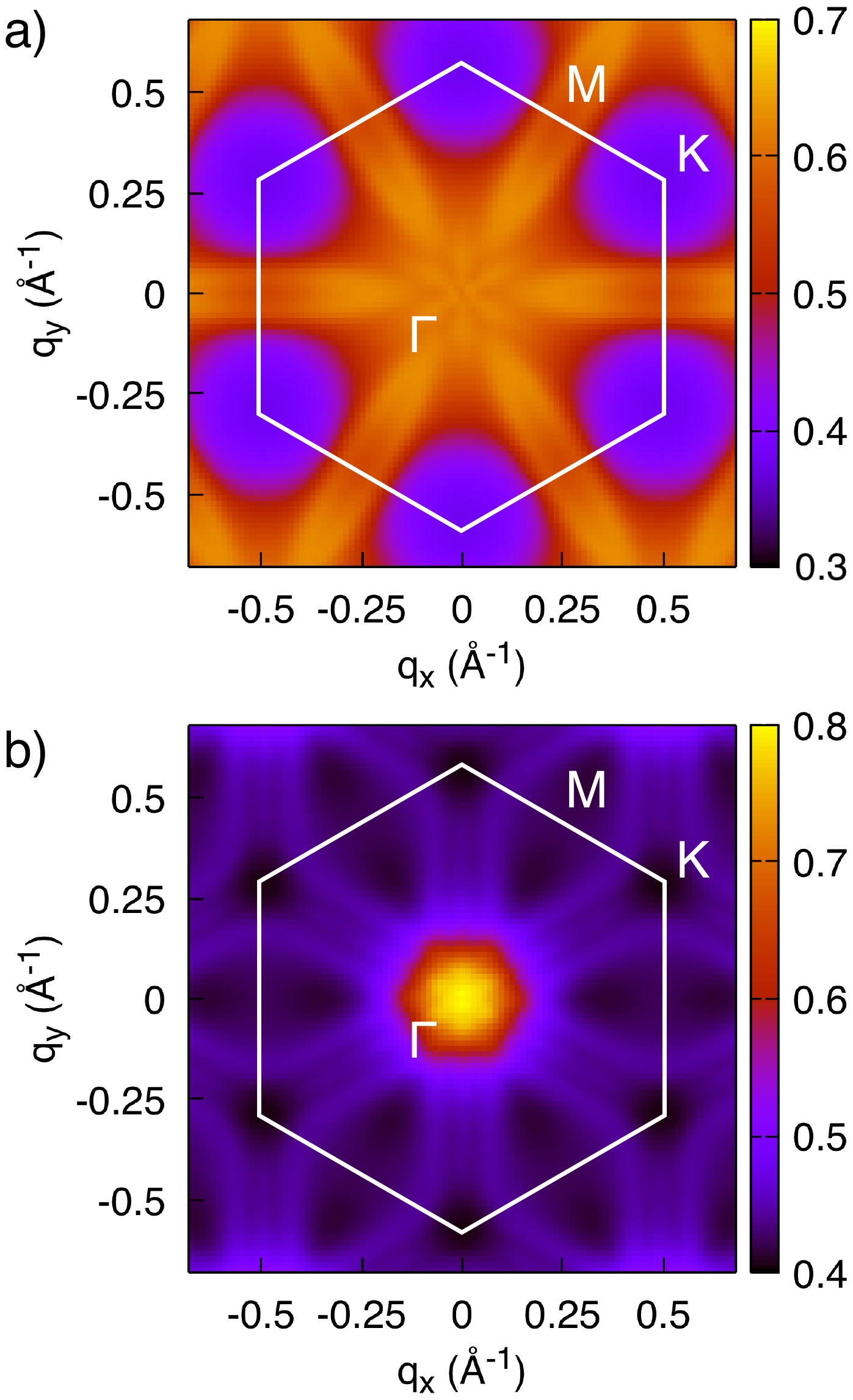}
\caption{(Color online) Distribution of the bare particle-hole susceptibility $\chi_0^{ph}({\bf q}) = \frac{1}{2}\sum_{l_1l_2}[\chi_0^{ph}({\bf q})]_{l_1l_2}^{l_1l_2}$ (at the HSE level) at the a) type-I vHs
and b) type-II vHs, respectively.}
\label{fig3}
\end{figure}

{\it Superconducting instabilities --} In order to account for
electronic interactions, we consider the onsite Hubbard model on the 2D hexagonal lattice,
with the non-interacting single particle Hamiltonian given
by~\eqref{eq1}, and $U$ parametrizing the Hubbard coupling
strength. The full
Hamiltonian reads
\begin{eqnarray}
\label{eq3}
H^{\text{J}} = H_0^{\text{J}} + U\sum_i n_{i\uparrow}n_{i\downarrow} - \mu\sum_{i,\sigma} n_{i,\sigma},
\end{eqnarray}
where $\mu$ is the chemical potential tuned to access the two vH
regimes. The combination of Fermi level density of states and finite
$U$ triggers superconducting instabilities, which we analyze in the following. 
At the
simplified RPA level, where the electronic two-particle vertex function is replaced by the bare interaction $U$, an effective
attractive interaction can emerge through the exchange of charge and magnetic fluctuations. These are governed by
the respective charge and magnetic susceptibilities $\chi_{c/m}^{ph}({\bf q})=[1 \pm U\chi_0^{ph}({\bf q})]^{-1}\chi_0^{ph}({\bf q})$.
The zero-frequency component of the bare susceptibility matrix in
the particle-hole channel is defined as
\begin{eqnarray}
\label{eq4}
[\chi_0^{ph}({\bf q})]_{l_2l_4}^{l_1l_3} &=& -\frac{1}{N_k}\sum_{{\bf k},nm} a_n^{l_2}({\bf k})a_n^{l_4*}({\bf k})a_m^{l_3}({\bf k}+{\bf q}) \nonumber \\
&&a_m^{l_1*}({\bf k}+{\bf q})\frac{f(\varepsilon_{n{\bf k}}) - f(\varepsilon_{m{\bf k}+{\bf q}})}{\varepsilon_{n{\bf k}} - \varepsilon_{m{\bf k}+{\bf q}}},
\end{eqnarray}
where $l_i = 1,2$ is the sublattice index and $a_n^{l_i}({\bf k})$ is the $l_i$th component of the $n$th eigenvector.
This quantity reveals the distribution of momentum transfer ${\bf q}$
implied by spin and charge fluctuations.

In Fig. \ref{fig3}a we show the susceptibility $\chi_0^{ph}({\bf q})$
at the type-I vHs. Significant intensity close to the {\bf M} point suggests dominant antiferromagnetic
spin fluctuations in the system, and is a consequence of the high degree of nesting of the Fermi surface at the
type-I vHs, as evident from Fig. \ref{fig2}b. Such a setting for spin fluctuations has
been found to trigger an even-parity chiral superconducting instability \cite{Graphene3RG,GrapheneFRG2,GrapheneFRG}. At the type-II vHs, on the other hand,
there is no Fermi surface nesting, and $\chi_0^{ph}({\bf q})$ is
solely peaked around the $\bf \Gamma$ point, suggesting that in this case, the dominant magnetic
fluctuations involve a long range modulation (i.e., ferromagnetic
fluctuations in the limit ${\bf q}\rightarrow 0$). General arguments
based on analytical weak coupling renormalization group applied to saddle points located
not at TRIM, close to type-II vHs, point to a
spin-triplet odd-parity superconducting state~\cite{Yao,vHsII}.

In order to provide a most substantiated analysis of the
superconducting instabilities, we apply a combined effort of
RPA and FRG. For the problem at hand, we find that all
approaches we have used reach the same conclusion on the nature of the
superconducting state. Since the FRG tracks vertex corrections and treats all instability channels on equal
footing, we choose to discuss the FRG results in the main text, and defer the confirming
evidence from RPA to the supplement~\cite{Supp}.
Within FRG, we formulate a set of coupled integro-differential
equations which describes a two-particle vertex flow equation $V_{\Lambda}$ where
the temperature flow parameter $\Lambda$ corresponds to the cutoff
parameter that evolves from high energies
towards the Fermi level~\cite{MetznerRMP,PlattReview}. Within the patch-FRG we employ here, the
two-particle vertex is projected to the
Fermi level, and discretized into $N=96$ patches.
The initial condition for the $96^3\sim8.8\times 10^5$-dimensional system of
integro-differential equations is given by the many-body interaction $U$. In
order to further improve the numerical performance, we consider the
HSE ab-initio band structure without SOC in order to exploit full
SU(2) symmetry, and obtain the spin triplet and singlet sectors by
vertex antisymmetrization and symmetrization, respectively. This
approximation is justified for the case of jacutingaite. As already
hinted at in Fig.~\ref{fig2}a and carefully checked by us, the SOC term predominantly serves to
open a band gap, but hardly affects the Fermi surface dispersion and
eigenstates at the electron and hole doped van Hove levels. A minor
difference is given for the precise location of the saddle points
along the {\bf K}$-${\bf M} lines for the type-II vHs, or the degree
of warping for the type-I vHs. Facing the choice between enhanced
radial resolution via more patches and tracking those minor
differences in terms of SOC-inclusive Fermiology, we find it
preferable to keep maximal radial resolution.

Within FRG, the renormalized
interaction $V_{\Lambda}$ 
starts to diverge in some channel as the infrared cutoff $\Lambda$ approaches the
Fermi surface; this marks the onset of a leading instability, which we
subsequently analyze within mean field theory~\cite{Rohe2007}. The
FRG
procedure adjusted to Fermi surface instabilities of interacting fermions  allows for an equal treatment of all
possible 2-particle instabilities, which is
an immediate advantage in comparison to RPA
where the procedure is constrained to only a single 2-particle channel of
interest, such as particle-particle or particle-hole. While the precise validity range of FRG in terms of
interaction strength still cannot be rigorously specified, it
provides numerical guidance to model
interacting electron systems at intermediate coupling.
\begin{figure}[!t]
\centering
\includegraphics[width=\columnwidth,angle=0,clip=true]{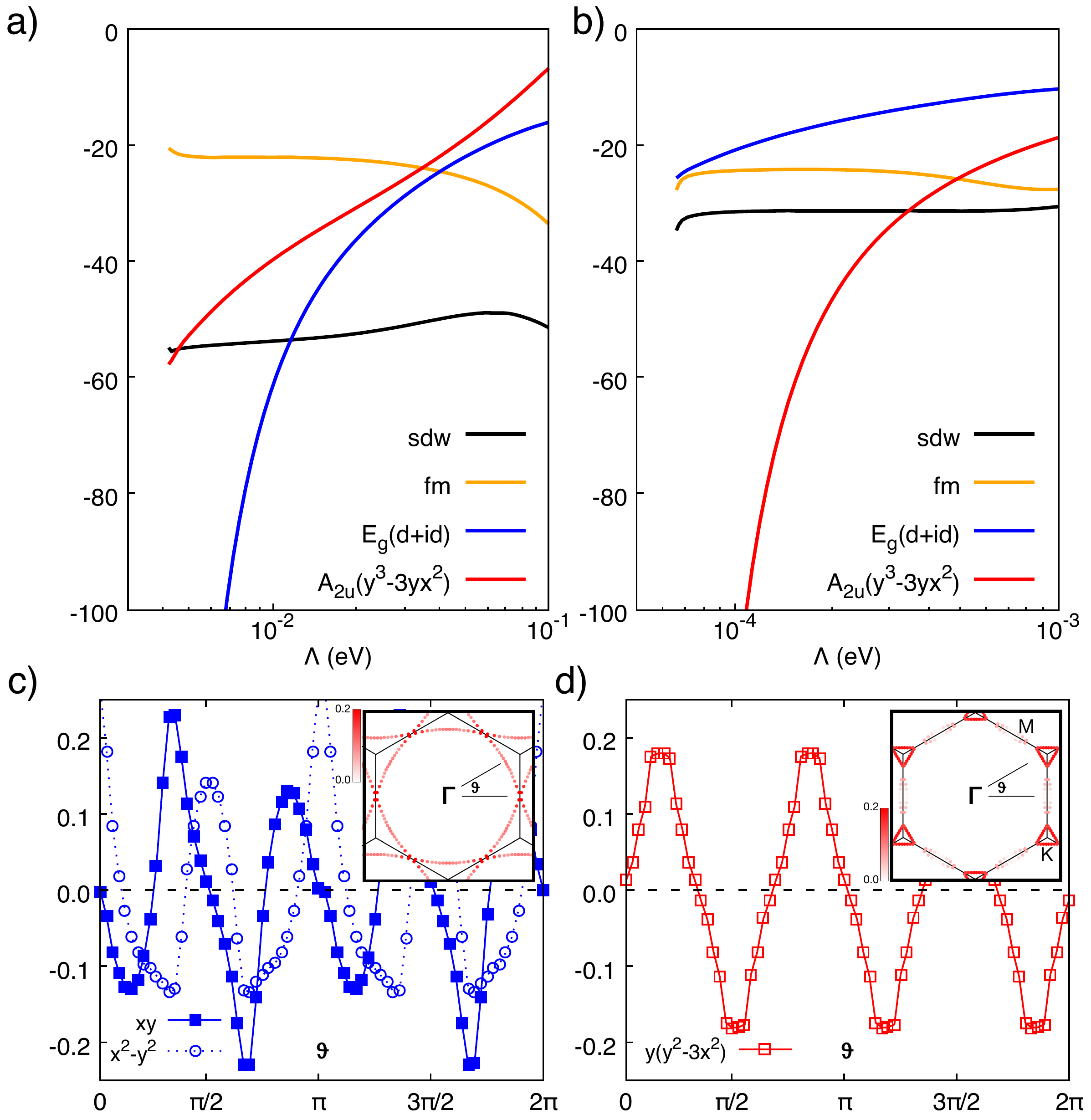}
\caption{(Color online) FRG flow for monolayer jacutingaite at the a)
  type-I vHs and b) type-II vHs ($U = 2.5$).  
c) and d) show the form factors along the Fermi surface
for the leading superconducting instabilities found in a) and b), respectively.
The insets in c) and d) report the gap function along the Fermi
surface from a subsequent treatment of the leading FRG instabilities in mean field theory.}
\label{fig4}
\end{figure}

Fig. \ref{fig4}a shows that when the chemical potential locates nearby
the type-I vHs, the effective interaction $V_{\Lambda}$ diverges in the even-parity spin-singlet superconducting
channel.
Its two degenerate order parameters $d_{xy}$ and $d_{x^2-y^2}$ transform as the
two-dimensional $E_g$ irreducible representation of $D_{3d}$, the point group of the buckled
honeycomb structure. The gap functions $\Delta_{d_{xy}}$ and $\Delta_{d_{x^2-y^2}}$ both have
line nodes along the Fermi surface (form factors depicted in
Fig. \ref{fig4}c). As evident from a subsequent mean field treatment,
the system gains condensation energy below the instability level by removing the
nodes via complex superposition $d_{xy} \pm id_{x^2-y^2}$ (inset of Fig. \ref{fig4}c), a
manifestation of spontaneous time reversal symmetry breaking.
As a subleading pairing channel, $f-$wave pairing
emerges, which is
also observed within RPA~\cite{Supp}.

When the chemical potential is shifted to the type-II
vHs, the Fermi surface is dominated by ferromagnetic fluctuations,
which favors a superconducting
instability in the spin-triplet sector (Fig. \ref{fig4}b). The form
factor of the leading instability transforms according
to the one-dimensional $A_{2u}$ irreducible representation of $D_{3d}$, {\it i.e.} the superconducting
state resides in the $f_{y(3x^2-y^2)}-$wave state (Fig.~\ref{fig4}d).
Again, RPA calculations are consistent with FRG in this setting~\cite{Supp}.


{\it Conclusions --} We have identified monolayer jacutingaite as a
promising host not only for a quantum spin Hall phase at pristine
filling~\cite{Marrazzo}, but also for unconventional superconductivity at van Hove
filling for electron and hole doping. The type-I vHs is reached upon doping by 0.39 holes, which corresponds to about
9.8\% hole doping. For the type-II vHs, even only 4.0\% electron doping ($\sim$0.16 electrons)
is needed, a value that in principle may even already be achieved
by electrolytic gating. In addition to the interest generated due to its
inherently exotic nature, the high experimental feasibility
of possibly accomplishing a type-II van Hove level without chemical doping is a
highly appealing feature of monolayer jacutingaite.
Note that, for instance, several attempts were made to dope graphene to
the vHs point by Ca and K adsorbates \cite{DopedGraphene}. Notwithstanding the
efforts, so far no evidence of superconductivity was
reported. Conversely, the rather small amount of doping needed to reach the type-II vHs renders jacutingaite a
promising material candidate to realistically
achieve unconventional superconductivity in a doped quantum spin Hall insulator.


We thank A.~Marrazzo and G.~Sangiovanni for discussions. This work was supported by the DFG through
SFB1170 "Tocotronics" (project B04) and by
ERC-StG-336012-Thomale-TOPOLECTRICS. We gratefully acknowledge the
Gauss Centre for Supercomputing e.V. (www.gauss-centre.eu) for funding
this project by providing computing time on the GCS Supercomputer
SuperMUC at Leibniz Supercomputing Centre (www.lrz.de). 

\bibliographystyle{prsty}
\bibliography{biblio}


\onecolumngrid
\newpage
\begin{center}
\textbf{\large Supplemental material for "Unconventional superconductivity in a doped quantum spin Hall insulator"}
\end{center}
\setcounter{equation}{0}
\setcounter{figure}{0}
\setcounter{table}{0}
\setcounter{page}{1}
\makeatletter
\renewcommand{\theequation}{S\arabic{equation}}
\renewcommand{\thetable}{S\Roman{table}}
\renewcommand{\thefigure}{S\arabic{figure}}

\section{Type I and type II van Hove singularities from long range hopping}
Neglecting spin-orbit coupling and considering long range hopping
terms, the Hamiltonian matrix of our effective model for the buckled honeycomb lattice is given by
\begin{eqnarray}
H_0^{\text{J}}=
\left(\begin{array}{cc}  H_{AA} &H_{AB} \\ H_{BA}  & H_{BB} \\  \end{array}\right),\label{supp1}
\end{eqnarray}
where $A/B$ refers to the sublattice index. The matrix elements of the Hamiltonian with up to 4th nearest neighbor hopping are
\begin{eqnarray}
H_{AB}&=&t_1(e^{\frac{ik_xa_0}{\sqrt{3}}}+2e^{-\frac{ik_xa_0}{2\sqrt{3}}}
          \cos \frac{k_y a_0}{2})+t_3(e^{\frac{-2ik_xa_0}{\sqrt{3}}}+2e^{\frac{ik_xa_0}{\sqrt{3}}}\cos k_y a_0)\nonumber\\
&&+2t_4(e^{\frac{5ik_xa_0}{2\sqrt{3}}}\cos \frac{1}{2}k_ya_0+e^{\frac{-2ik_xa_0}{\sqrt{3}}}\cos k_y a_0+e^{\frac{-ik_xa_0}{2\sqrt{3}}}\cos\frac{3}{2} k_y a_0),\\
H_{BA}&=&H^*_{AB},\\
H_{AA}&=&H_{BB}=t_2(2\cos k_ya_0+4\cos \frac{\sqrt{3}}{2}k_xa_0\cos \frac{1}{2}k_ya_0).
\end{eqnarray}
Here, $t_2$ denotes an intra-sublattice and $t_{1,3,4}$ inter-sublattice hopping integrals. As the $t_2$ term breaks chiral
symmetry, the spectrum is asymmetric with respect to half filling.
If one were to consider only
inter-sublattice hopping, the energy spectrum would exhibit a chiral symmetry;
the $t_2$ term, however, is diagonal in the
sublattice space, and hence imposes a $k$ dependent energy shift for
the spectrum. As a consequence of such a shift, the spectrum becomes
asymmetric with respect to half filling. When only the NN hopping is
included, the Fermi surface develops a perfect
hexagonal shape when the Fermi level is exactly at the van Hove singularity point. The
saddle point is located at the time reversal invariant momentum, and the
corresponding vHs point is dubbed type-I. If the saddle point is
not at the time reversal invariant momentum, the corresponding vHs point is
called type-II.

\begin{figure}[tb]
\centerline{\includegraphics[height=4 cm]{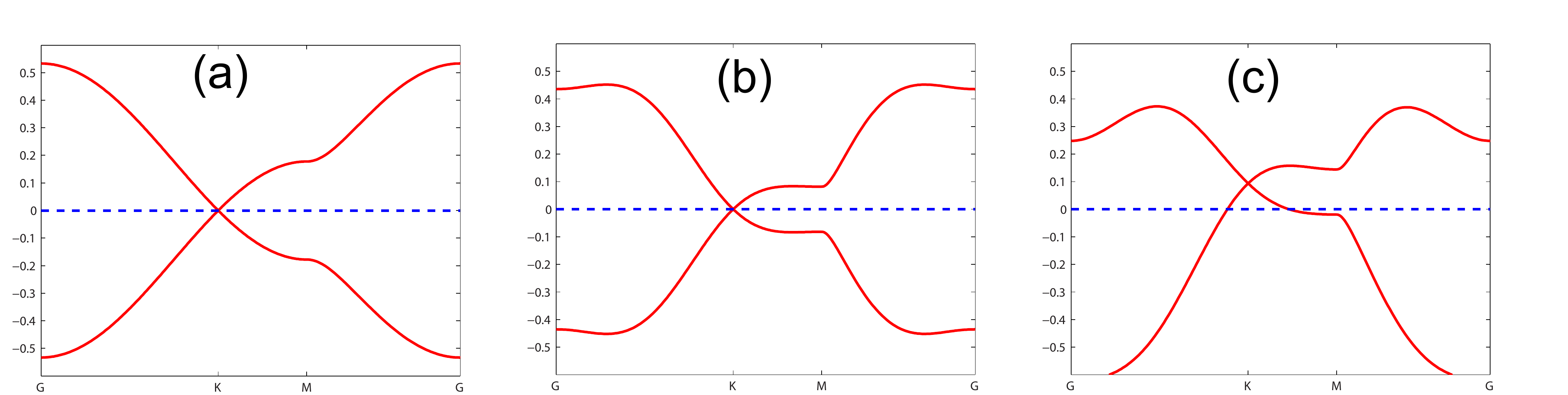}}
\caption{(color online) Band structures related to~\eqref{supp1}. (a) with only NN hopping $t_1$, (b) with hopping $t_1$, $t_3$ and $t_4$, (d) with all hoppings up to 4th NN.  \label{band} }
\end{figure}

Along the K-M line, i.e. for $k_x = \frac{2\pi}{\sqrt{3}a_0}$, the elements of the Hamiltonian are
\begin{eqnarray}
H_{AB}&=&ht_1+ht_3+ht_4,\\
H_{AA}&=&H_{BB}=ht_2,\\
ht_1&=&t_1e^{\frac{2i\pi}{3}}f_1(k)=t_1e^{\frac{2i\pi}{3}}(1-2\cos \frac{k_ya_0}{2}),\\
ht_2&=&t_2f_2(k)=t_2(2\cos k_ya_0-4\cos \frac{1}{2}k_ya_0),\\
ht_3&=&t_3e^{\frac{2i\pi}{3}}f_3(k)=t_3e^{\frac{2i\pi}{3}}(2\cos k_ya_0+1),\\
ht_4&=&t_4e^{-\frac{i\pi}{3}}f_4(k)=t_4e^{-\frac{i\pi}{3}}(\cos\frac{k_ya_0}{2}-cosk_ya_0+\cos
        \frac{3}{2}k_ya_0).
\end{eqnarray}
At the M $(\frac{2\pi}{\sqrt{3}a_0},0)$ point, we have
$H_{AB}=(-t_1+3t_3-2t_4)e^{\frac{2i\pi}{3}}$ and $H_{AA/BB}=-2t_2$.
At the K$(\frac{2\pi}{\sqrt{3}a_0},\frac{2\pi}{3a_0})$ point instead,
$H_{AB}=0$ and $H_{AA/BB}=-3t_2$. In a monolayer of jacutingaite,
$t_{1,3}>0$ and $t_4<0$, such that $t_3/t_4$ will partly compensate $t_1$ in
$H_{AB}$, generating a flat band along the KM line (as shown in
Fig.\ref{band}(b)). With an eventual inclusion of the $t_2$ term, we end up with a
type-II VHS point in the conduction band and a type-I VHS point in the valence
band, as shown in Fig.\ref{band}(c).

%
%

%

\section{First-principles DFT calculations}

First-principles calculations were performed by using the
projector augmented plane wave method as implemented in the Vienna
Ab-initio Simulation Package (VASP) [1]. We adopt both the PBE [2] as well as the
Heyd-Scuseria-Ernzerhof (HSE) hybrid functional [3] for the
exchange-correlation part, while SOC has been included
self-consistently in a scalar-relativistic scheme. We use a plane
wave cutoff of 600 eV on a 12$\times$12$\times$1 Monkhorst-Pack $k$-point mesh.
The projection onto maximally localized wannier functions was achieved via the
WANNIER90 package [4].

\begin{figure}[tb]
\centerline{\includegraphics[height=6 cm]{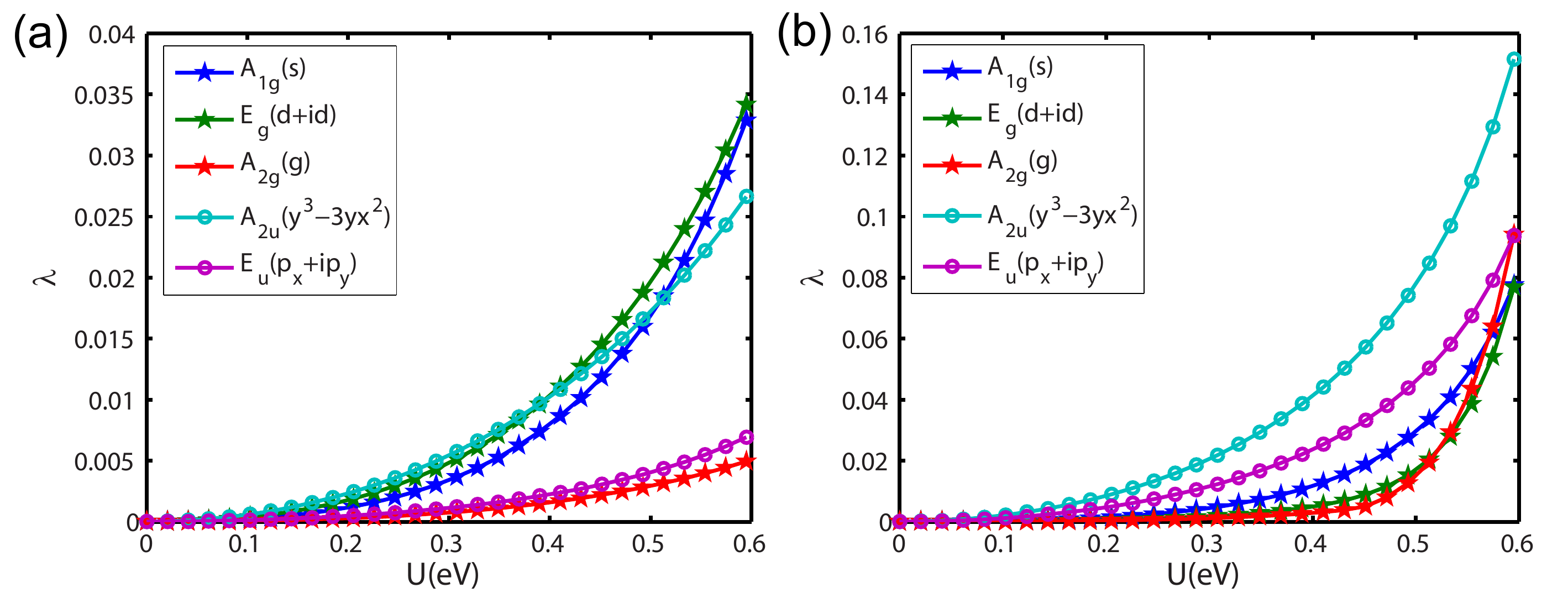}}
\caption{RPA pairing strengths as a function of $U$ at (a) type-I van Hove
  filling and (b) type-II van Hove filling.  \label{T12} }
\end{figure}

\begin{figure}[tb]
\centerline{\includegraphics[height=4 cm]{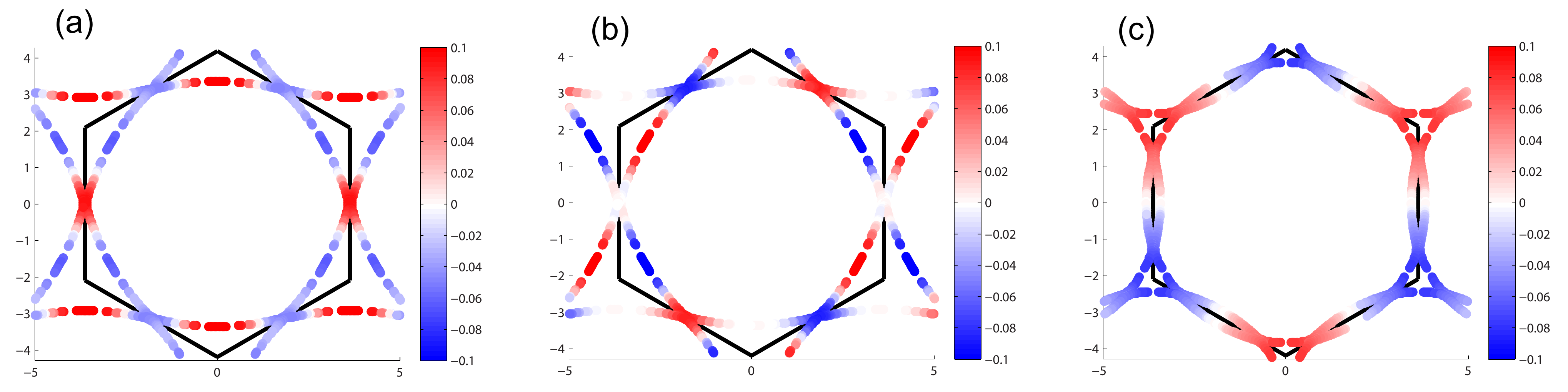}}
\caption{RPA gap functions of the most dominant pairing states at
  type-I and type-II van Hove filling: (a) $E_{2g}(d_{x^2-y^2}$), (b) $E_{2g}(d_{xy}$); (c) $A_{2u}(f_{y^3-3yx^2})$.  \label{gap} }
\end{figure}

\section{RPA calculations}
 We perform RPA calculations by following those described in
 Refs. [5, 6]. The pairing
 strengths of the dominant pairing states as a function of interaction
 at type-I and type-II VHS points are displayed in Fig.\ref{T12}(a) and
 (b), respectively. At the type-I vHs point, $d+id$ pairing state is the
 most dominant one for $0.4\leq U \leq 0.6$ eV due to the strong
 antiferromagnetic fluctuations from Fermi surface nesting, and competes strongly with the
 singlet $s-$wave and $f-$wave instabilities depending on the interaction strength.
 At the type-II vHs point, there is no Fermi surface nesting, and  $f_{y^3-3yx^2}$
 wave pairing state is the most dominant due to the ferromagnetic
 fluctuations. The gap functions of $E_{2g}$ and $A_{2u}$ states are
 shown in Fig.\ref{gap}. The findings are consistent with the FRG
 results presented and discussed in the main text.

\vspace{1cm}

[1] G. Kresse and J. Furthm\"{u}ller, Phys. Rev. B {\bf 54}, 11169 (1996).

[2] J. P. Perdew, K. Burke, and M. Ernzerhof, Phys. Rev. Lett. {\bf 77}, 3865 (1996).

[3] J. Heyd, G. E. Scuseria, and M. Ernzerhof, J. Chem. Phys. {\bf 121}, 1187 (2004).

[4] A. A. Mostofi et al., Comput. Phys. Commun. {\bf 178}, 685 (2008).

[5] S. Graser, T. A. Maier, P. J. Hirschfeld and D. J. Scalapino, New J. Phys., {\bf 11}, 025016 (2009)

[6] N. E. Bickers, D. J. Scalapino and S. R. White, Phys. Rev. Lett., {\bf 62} 961 (1989)

\end{document}